\documentclass[twocolumn,aps,prl,superscriptaddress,floatfix]{revtex4-1}
\usepackage{graphicx}
\usepackage{amsmath}
\usepackage{tabularx}
\usepackage[caption=false]{subfig}
\usepackage{epstopdf}

\begin{document}
\title{Fermi Surface Reconstruction in Underdoped Cuprates: The Origin of Electron Pockets}

\author{Ilya Ivantsov}
\affiliation{Bogoliubov Laboratory of Theoretical Physics, Joint Institute for Nuclear Research, Dubna, Russia}
\affiliation{L.V.Kyrensky Institute of Physics, Siberian Branch of Russian Academy of Sciences, Krasnoyarsk, Russia}
\author{Alvaro Ferraz}
\affiliation{International Institute of Physics - UFRN, Department of Experimental and Theoretical Physics - UFRN, Natal, Brazil}
\author{Evgenii Kochetov}
\affiliation{Bogoliubov Laboratory of Theoretical Physics, Joint Institute for Nuclear Research, Dubna, Russia}
\affiliation{International Institute of Physics - UFRN, Department of Experimental and Theoretical Physics - UFRN, Natal, Brazil}

\begin{abstract}

A new phenomenological model is proposed to describe the evolution of the Fermi surface (FS) in a wide range of dopings.
It reproduces the key features of the cuprates in the underdoped phase above the superconducting temperature $T_c$.
It is shown that the explicit accounting of strong electron correlation in the framework of the $t-J$ model taken in complementary to the translational symmetry breaking induced by the charge density wave (CDW) gives rise to the Fermi surface reconstruction (FSR) into small electron pockets.
While the strong Coulomb repulsion leads to an emergence of the arc-like Fermi surface in the pseudogap (PG) phase, the Bragg reflection on the boundaries of the reduced Brillouin zone (BZ) opens up a possibility to close the quasiparticle orbits.
Direct calculation of the FS properties allows us to unveil the scenario of the experimentally observed transition to the CDW phase that sets in at the doping level $\delta\approx0.08$ and is accompanied by a divergence of the carriers effective mass and the sign reversals of the Hall and Seebeck coefficients.

\end{abstract}

\maketitle

\paragraph{Introduction.} The discovery of quantum oscillations in the underdoped $\mathrm{YBa_2Cu_3O_{6+\delta}}$ (YBCO) in a high magnetic field\cite{DL2007,Yelland2007,Sebastian2008} revealed among other features a reconstruction of the Fermi surface into small pockets.
Further measurements\cite{Badoux2016_Nature} also established that the area of the FS is proportional to the doping value $\delta$ instead of the $1+\delta$ as predicted by the conventional Fermi liquid (FL) theory.
The sign changes of both Hall\cite{LeBoeuf2007} and Seebeck\cite{Chang2010} coefficients in the doping range of $0.08<\delta<0.16$ indicates that, at a low temperature, electron-like pockets emerge due to the translation symmetry breaking that takes place in the ground state of the non-superconducting phase.
The subsequent detection of quantum oscillations in $\mathrm{YBa_2Cu_4O_8}$\cite{Yelland2008,Banquara2008}, $\mathrm{HgBa_2CuO_{4+\delta}}$\cite{Barisic2013} and other compounds\cite{Suzuki2002,Badoux2016,Laliberte2011,Noda1999,Adachi2001} confirms that such a FSR is a generic property of the underdoped cuprates.

All those compounds reveal the presence of the CDW modulation right at the very doping level where the FSR has been detected.
It was shown that, for YBCO, the incommensurate CDW modulation vectors lie in the $\mathrm{CuO}$ plane and they take the form $\mathbf{Q}_x=2\pi (Q,0)$ and $\mathbf{Q}_y=2\pi (0,Q)$ with $Q\approx 0.31$\cite{Wu2011,Wu2013,Blanco2014,Tabis2014}.
This CDW ordering with a period varying from 3 to 5 lattice spacings has also been detected by scanning tunneling microscopy (STM) and x-ray diffraction (XRD) studies in underdoped cuprates \cite{Tranquada1995,Fink2011,Ghiringhelli2012,Achkar2012,LeBoeuf2013,Blackburn2013,Hoffman2002,Kohsaka2007,Kivelson2003,Vershinin2004}.
Those experimental evidences strongly suggest that this charge ordering indeed is the leading cause for the observed FSR.
Also, a number of experiments revealed a predominant $d$-wave form factor of the CDW ordering \cite{Comin2014,Fujita2014,Forgan2015}.

The phenomenological calculations confirm that the electron-like pockets in the nodal regions can be obtained within the one band phenomenological tight-binding model by breaking translation symmetry by multiple-Q ordering\cite{HS2011}.
Extending that model to incorporate a long-range charge order with the $d$-form factor \cite{Allias2014} produces small hole-like pockets in addition to the existing electron pockets in agreement with the experimental measurements\cite{DL2015}.
However, the FSR is observed within a still mysterious PG phase which is believed to be driven by strong electron correlations.
While most models neglect electron-electron interaction or, at most, account them in the framework of mean field theory\cite{Maharaj2014,Sachdev2014} the proximity of PG and CDW phase demands the more accurate accounting of the strong electron interaction to reproduce the whole evolution of the FS.

\paragraph{Model and Method.}

\begin{figure}
\includegraphics[width=1\linewidth]{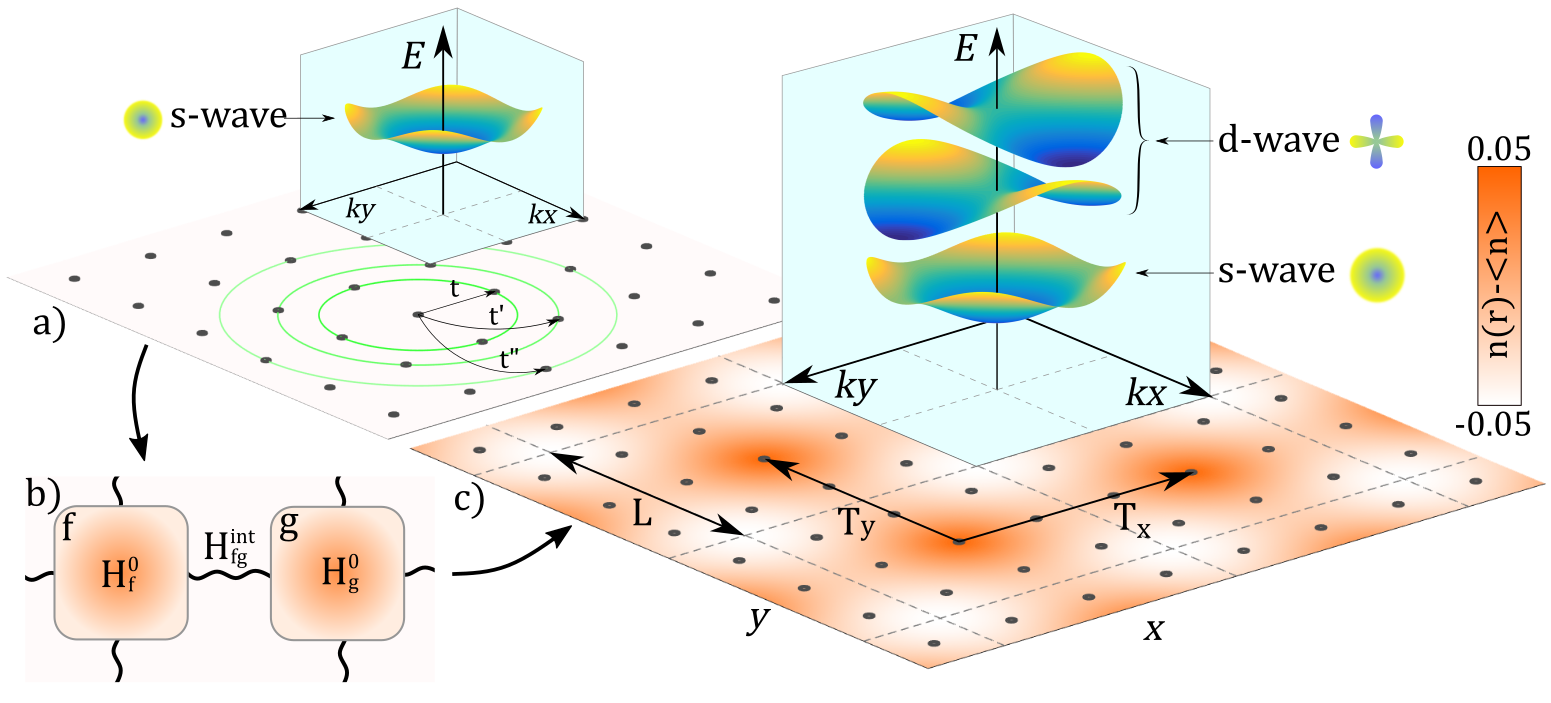}
\caption{{\bf (a)}: The picture of the original tight-binding model on a square lattice with the nearest $t$, next nearest $t'$ and next-next-nearest $t''$ neighbours hopping.
The inset schematically represents the $s$-wave hopping mode corresponding to the well known dispersion relation $\varepsilon(\mathbf{k})=2t(\mathrm{cos}(k_x)+\mathrm{cos}(k_y))+4t'\mathrm{cos}(k_x)\mathrm{cos}(k_y)+2t''(\mathrm{cos}(2k_x)+\mathrm{cos}(2k_y))$.
{\bf (b)}: A schematic representation of the Hamiltonian in the superlattice representation.
{\bf (c)}: The superlattice consisting of the $3\times 3$ clusters.
The spatial distribution of the charge density projected on the original lattice corresponds to the colorbar.
The inset schematically represents the modes of the intercluster interaction.}
\label{Fig1}
\end{figure}

We start with a tight-binding model on a square lattice (Fig.\ref{Fig1}(a)) where the translations are described by two unit vectors $\mathbf{E}_x=(1,0)$ and $\mathbf{E}_y=(0,1)$ (from now on we put the lattice spacing $a$ equal to $1$).
We then break this symmetry modulo the translations generated by $\mathbf{T}_x=(L,0)$ and $\mathbf{T}_y=(0,L)$ where $L$ is a positive integer, the wavevectors corresponding to these new translations being then $\mathbf{Q}_x=(\frac{2\pi}{L},0)$ and $\mathbf{Q}_y=(0,\frac{2\pi}{L})$.
In this way, the translation symmetry is persevered only along the superlattice of the basic clusters with size $L\times L$.
Since the XRD experiments yield the value of $L\approx 3.2$ (in YBCO)\cite{Blanco2014,Tabis2014}, we take the closest integer value of $L=3$.
In this way, we consider the $3\times 3$ clusters as the new "unit cells", with an appropriate multiband internal structure.

Explicitly, we employ the Hamiltonian in the superlattice representation:

\begin{equation}
H=\sum_{fg}H^{\mathrm{int}}_{fg}+\sum_fH^0_f
\mathrm{,}\quad
H^{\mathrm{int}}_{fg}=\sum_{ab\sigma}t^{fg}_{ab}c_{fa\sigma}^{\dagger}c_{gb\sigma}
\label{eq2}
\end{equation}
where $f$ and $g$ stand for the cluster indices and $a,b$ denote the sites inside the cluster, $t^{fg}_{ab}$ is the hopping amplitude and $c_{fa\sigma}^{\dagger}$ is the electron creation operator at the $a$-th site of cluster $f$ with the spin projection $\sigma$.
While the $H_{fg}^{\mathrm{int}}$ defines the general form of intercluster interaction the $H_f^0$ term corresponds to the intracluster contribution determining the inner structure of the clusters and, consequently, the specific form of intercluster hopping and its relation to the clusters eigenstates.
In case of a the purely tight-binding model, $H^0_f=H^{\mathrm{int}}_{ff}$, the resulting FS is a large FL-like cylinder.
In our approach, we modify the intracluster part by adding a strong on-site Coulomb repulsion $U\sum_a n_{fa\uparrow}n_{fa\downarrow}$ to take the $H^0_f$ in the $U$-large limit in the form of the $t-J$ model:
\begin{equation}
H^0_f=\sum_{ab\sigma}(t_{ab}^{ff}-\mu\delta_{ab})\tilde{c}_{fa\sigma}^{\dagger}\tilde{c}_{fb\sigma}+J\sum_{ab}\mathbf{S}_{fa}\cdot \mathbf{S}_{fb}.
\label{eq3}
\end{equation}
Here the spin exchange coupling $J=4t^2/U$, the constrained electron operators $\tilde{c}_{a\sigma}=c_{a\sigma}(1-n_{a-\sigma})$, and $\mathbf{S}_{fa}$ is the electron spin operator.
The next step is a numerical exact diagonalization of the $H^0_f$ for each cluster which allows us to rewrite the Hamiltonian (Eq.\ref{eq2}) in the basis of the clusters eigenstates and calculate the electron spectral function $A(\mathbf{k},\varepsilon)$ for the entire lattice by using the
so-called cluster perturbation theory (CPT) \cite{Senechal2000,Senechal2002,Maier2005}.

Two points are to be noted at this stage.
First, due to the finite cluster size a charge distribution inside the cluster becomes inhomogeneously modulated by the CDW with the wavevectors $\mathbf{Q}_x$ and $\mathbf{Q}_y$ as depicted in Fig.\ref{Fig1}(c).
This fact allows us to reproduce the experimentally observed (commensurate) bidirectional charge density wave\cite{Tabis2014}.
Second, since the phenomenology based on the tight-binding models contains only the $s$-form hopping, the $d$-form factor ordering can, in these models, be obtained only by explicit adding a corresponding long-range ordering term \cite{Allias2014}.
In our approach, the intercluster interaction is determined through the eigenstates of the clusters and Coulomb repulsion leads to specific short-range correlations responsible for the appearance of $d$-wave mode in the intercluster hopping term.
We thus avoid any predetermined ordering in the model so that the FSR appears as a natural result of {\it short-distance effects} which are explicitly taken into consideration, with a single phenomenological parameter being $Q = 1/3$.

\begin{figure}
\includegraphics[width=1\linewidth]{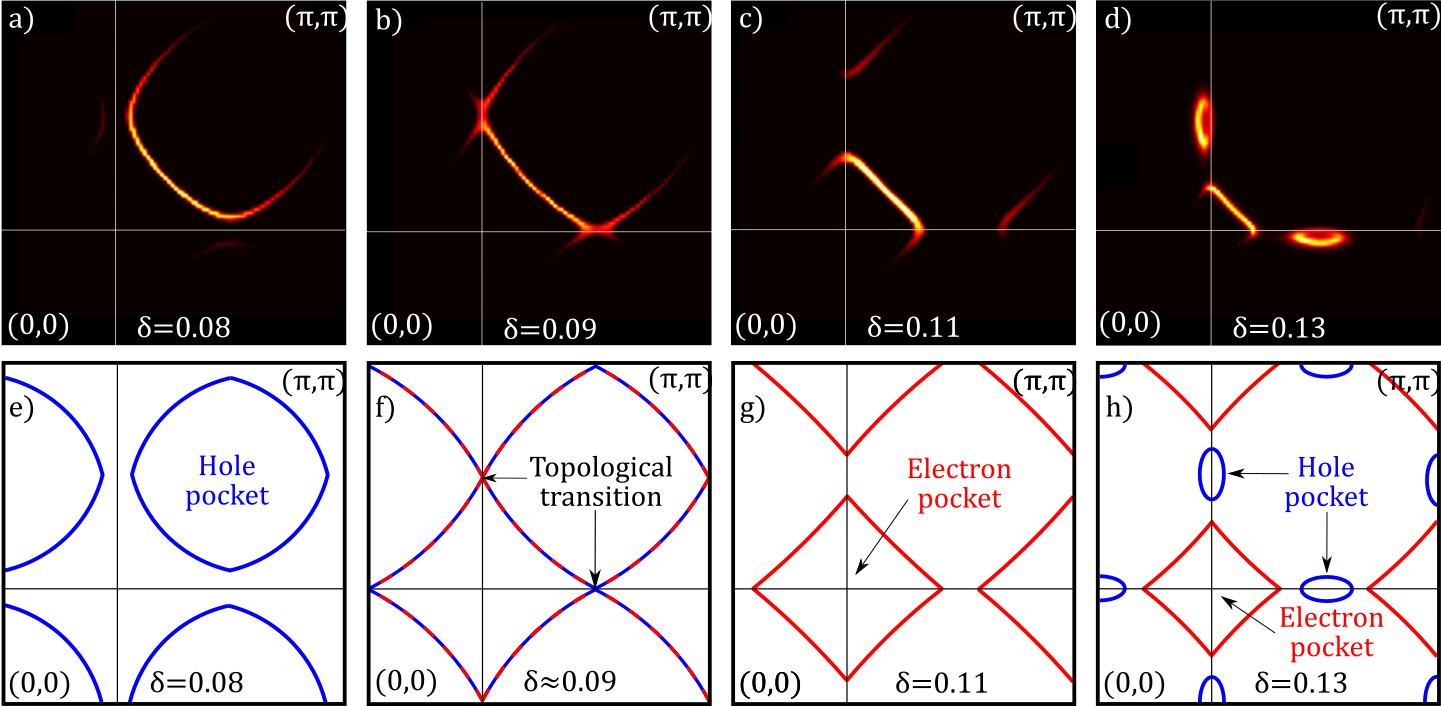}
\caption{The Fermi surfaces calculated for the different doping levels $\delta$ with ({\bf a-d}) and without ({\bf e-f}) taking into consideration the distribution of the spectral weight.
The solid lines $k_x=\pi/3$ and $k_y=\pi/3$ represent the Bragg "planes" in the reconstructed BZ.
The parameters of $t-J$ model used in the calculation are $t'=-0.27$, $t''=0.2$, $J=0.5$, temperature $T=10^{-4}$ (here and below the energy scale is in units of $t$).}
\label{Fig2}
\end{figure}

\begin{figure*}
\includegraphics[width=1\linewidth]{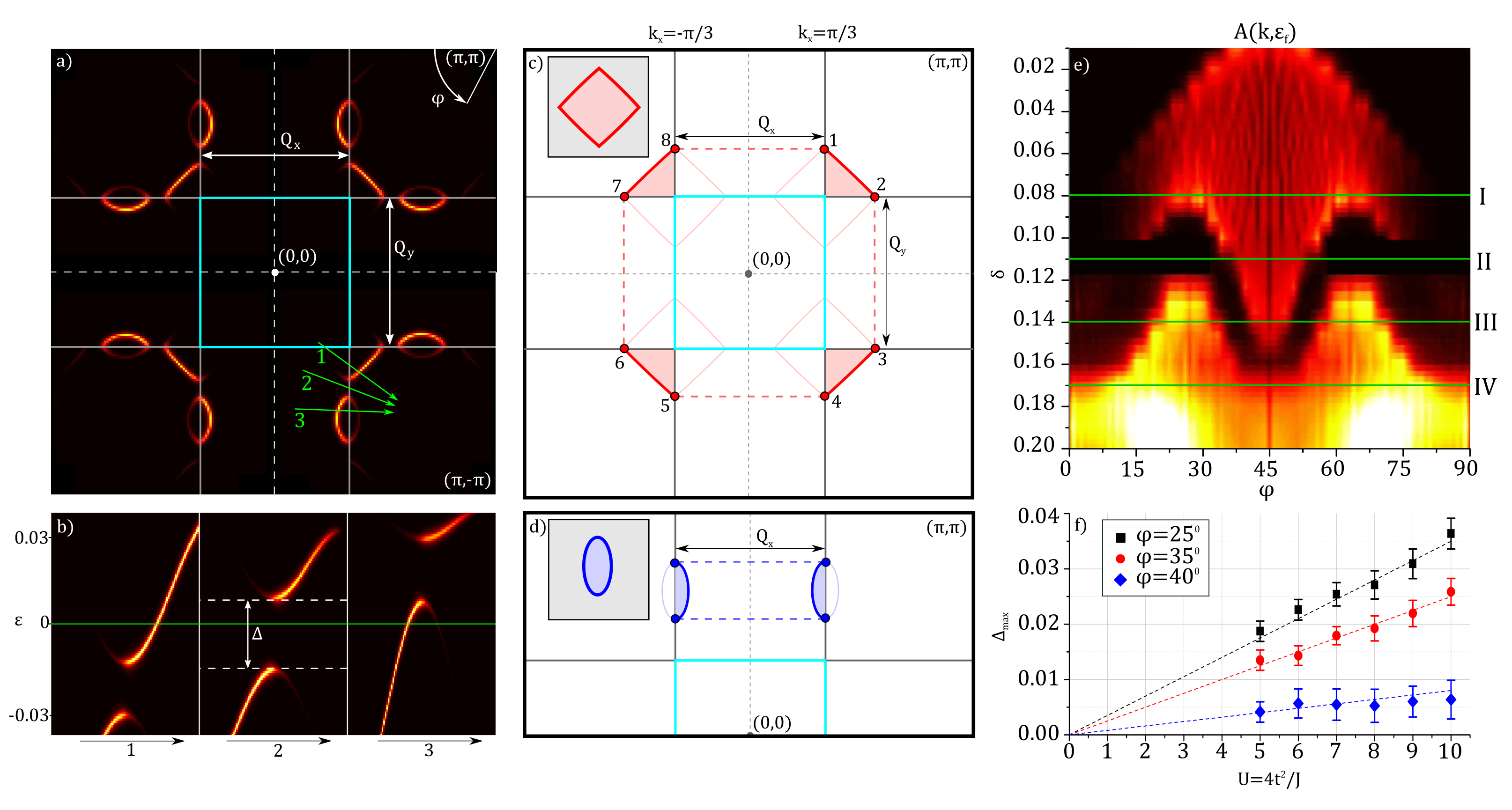}
\caption{{\bf (a)}: The FS calculated by the CPT method at the doping level $\delta=0.14$ and with the Lorentzian broadening value $\eta=0.01$.
The slices of the spectral function in the direction $1,2$ and $3$ (depicted in green) are presented in panel {\bf (b)}: The horizontal solid line denotes the Fermi level.
$\Delta$ denotes the value of the gap resulting from the band splitting.
{\bf (c)} Schematic representation of the electron pocket formation.
The reconstructed FS is depicted in the repeated BZ representation.
Enumerated dots correspond to the Bragg reflection points.
The sides of the electron pockets giving rise to the FS spectral weight are indicated in red and the area contributing to the full area of pocket is depicted by shading.
The inset schematically represent the resulted closed pocket.
{\bf (d)} Schematic representation of the formation of the hole pockets due to the Bragg reflection.
On the panels {\bf (a,c,d)} the first reconstructed Brillouin zone is depicted as the cyan square, the Bragg planes correspond to the solid lines $k_x, k_y=\pm\pi/3$, $Q_x$ and $Q_y$ correspond to the CDW modulation vectors.
{\bf (e)} The distribution of the FS spectral weights in the direction specified by the angle $\varphi$ (panel {\bf (a)}) with various doping levels calculated by the CPT method.
To each horizontal slice corresponds a doping $\delta$ denoted on the vertical axis.
The slices indicated by the Roman numerals correspond to the "phases" depicted in Fig.\ref{Fig4}.
The line III also corresponds to the FS depicted in panel {\bf (b)}.
{\bf (f)} The dependence of the value of the gap maximum on the Coulomb repulsion $U$.}
\label{Fig3}
\end{figure*}

\paragraph{Results.}

Our approach allows us to present a detailed scheme for the FSR in the whole PG phase ranging from a low to optimal doping.
Let us start with the low doping regime already outside the antiferromagnetic (AF) region, in which the experimental data reveal the absence of quantum oscillations\cite{LeBoeuf2011} which implies the absence of a FL regime and the FS appears to consist of nodal arcs as confirmed by the ARPES measurements\cite{Yang2011,Comin2014}.
The physics behind this arcs is still unclear, however.
Theoretical models\cite{YRZ2006} as well as numerical calculations\cite{IFK2017,Stanescu2006} suggest that the arcs occur due to the rearrangement of the spectral weight along the sides of a hole-like pocket.
Our calculation in the "pure" PG ($\delta<0.08$) phase confirm that the Fermi surface is made of hole-like pockets located in the nodal regions of the reduced Brillouin zone (BZ) that do not touch the boundaries of the reduced BZ as depicted in (Fig.\ref{Fig2}(e)).
The CPT calculation reveal that the spectral weight is non-zero only on one side of the pocket in such way that the FS simply reduces to the Fermi arc (Fig.\ref{Fig2}(a)).
In this way, our consideration strongly suggests that the vector connecting the tips of the arcs is distinctly different from the CDW modulation vector.

Despite the fact that the translational symmetry breaking is implemented "by hand" regardless of the doping, the obtained FS correctly emerges as the arcs in agreement with the ARPES measurement predicting the observed zero spectral weight in the antinodal region\cite{Shen2005}.
It is due to the fact, that the PG phase is driven by strong on-site Coulomb repulsion which leads to the arc-like structure in the underdoped phase {\it per se}\cite{Senechal2004}.
Therefore it turns out that the translational symmetry breaking does not seem to affect the structure of the FS in this "pure" PG regime.
This statement is supported by a number of numerical calculations carried out in the framework of cluster approaches (CPT\cite{Korshunov2007,Kohno2012,KNO2014}, C-DMFT\cite{Stanescu2006,Ferrero2009,Sakai2010}) for the different cluster sizes, whith the resulting FSs qualitatively corresponding to the FS arcs obtained in the present work.

However, in contrast with that, at larger dopings the translational symmetry breaking does affect the FS reconstruction\cite{HS2014,HS2015} in full agreement with experiment.
Namely, as doping increases, the FS undergoes a topology change (Fig.\ref{Fig2}(f)) at the doping level $\delta\approx 0.09$ that can be directly associated with the divergence of the carriers effective mass\cite{Sebastian2010,Ramshaw2015}.
At this doping, the FS starts to intersect the boundaries of the reduced BZ.
Moreover, at those points the vectors connecting the arc tips and the CDW modulation vectors coincide with each other. This signals the emergence of the CDW phase.
The electrons can now be scattered by the Bragg planes to form closed pockets in the momentum space as shown in Fig.\ref{Fig3}(c).
However, only one side of those pockets has a non-zeros spectral weight just like in the "pure" PG phase.
Explicitly, the "touching" points $(1,8), (2,3), (4,5)$ and $(6,7)$ can be identified as Bragg's reflection points that connect the quasiparticle states with same energy but with the momenta differ by the modulation vector $\mathbf{Q}$.
This results in the emergence of the effectively closed quasiparticle orbits composed of the arcs located in the nodal region.
These closed orbits give rise to the quantum oscillations seen in thermodynamic quantities in a high magnetic field.
Basically the same mechanism underlies the formation of the hole-like pockets as shown in Fig.\ref{Fig3}(d).
The emergence of hole pockets produced by the presence of $d$-wave form factor ordering is also unveiled in other experiments performed in cuprates\cite{Comin2014,Fujita2014,Forgan2015}.

Note also that the electron pockets simply originate from the reconstruction of the BZ that does not violate, in this case, the Luttinger theorem.
Therefore the physics dominated by electron pockets, in contrast with the "pure" arc phase, does not seem to display a non-Fermi liquid behavior.
In fact, the low-energy transport in the underdoped cuprates dominated by electron-like pockets indeed displays a conventional FL behaviour\cite{LeBoeuf2011,Proust2016,Grissonnanche2016,Sebastian2010_prb}.

\begin{figure}[!]
\begin{minipage}[h]{0.99\linewidth}
\center{(a)\\\includegraphics[width=1\linewidth]{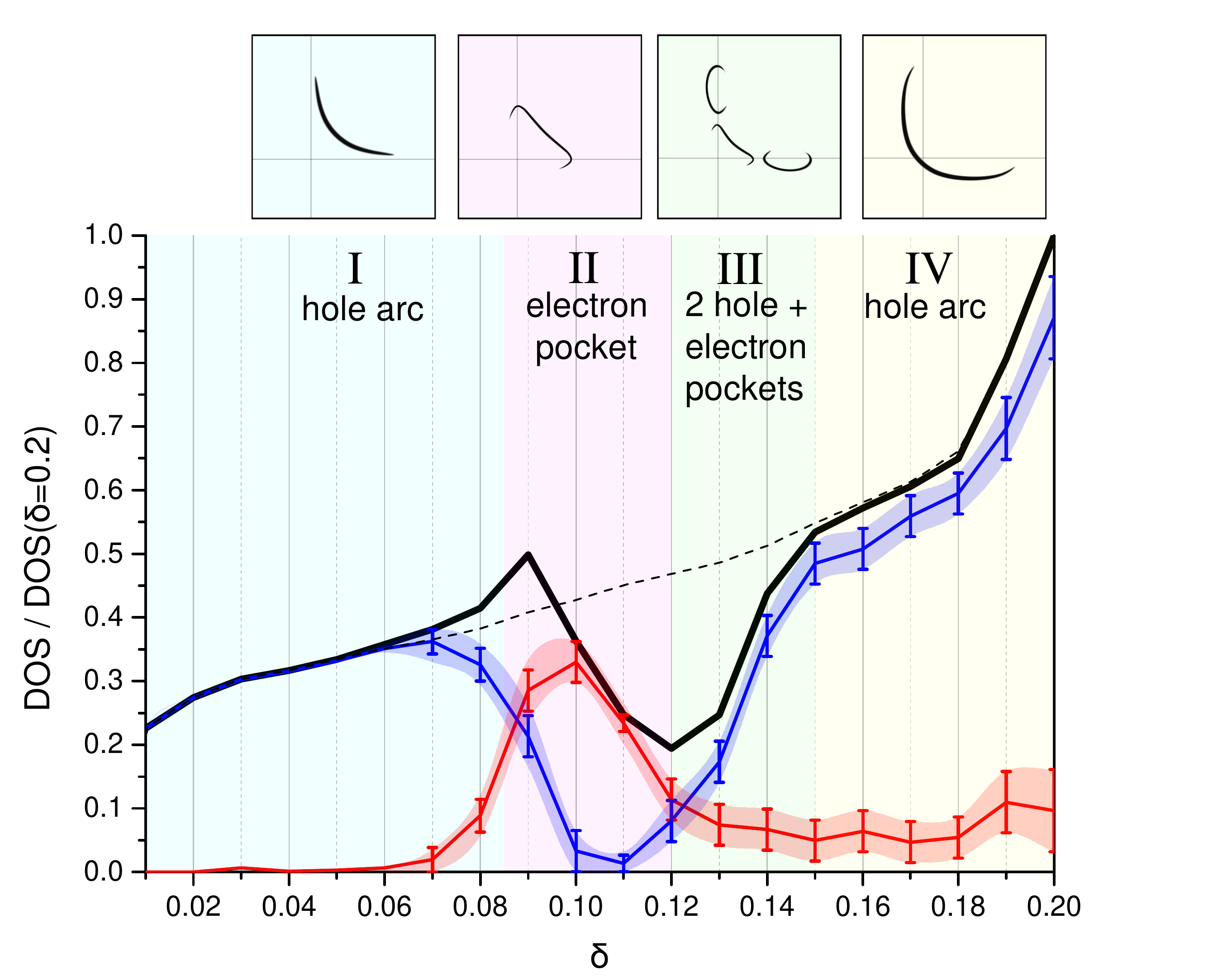}}
\end{minipage}
\vfill
\begin{minipage}[h]{0.99\linewidth}
\center{(b)\\\includegraphics[width=1\linewidth]{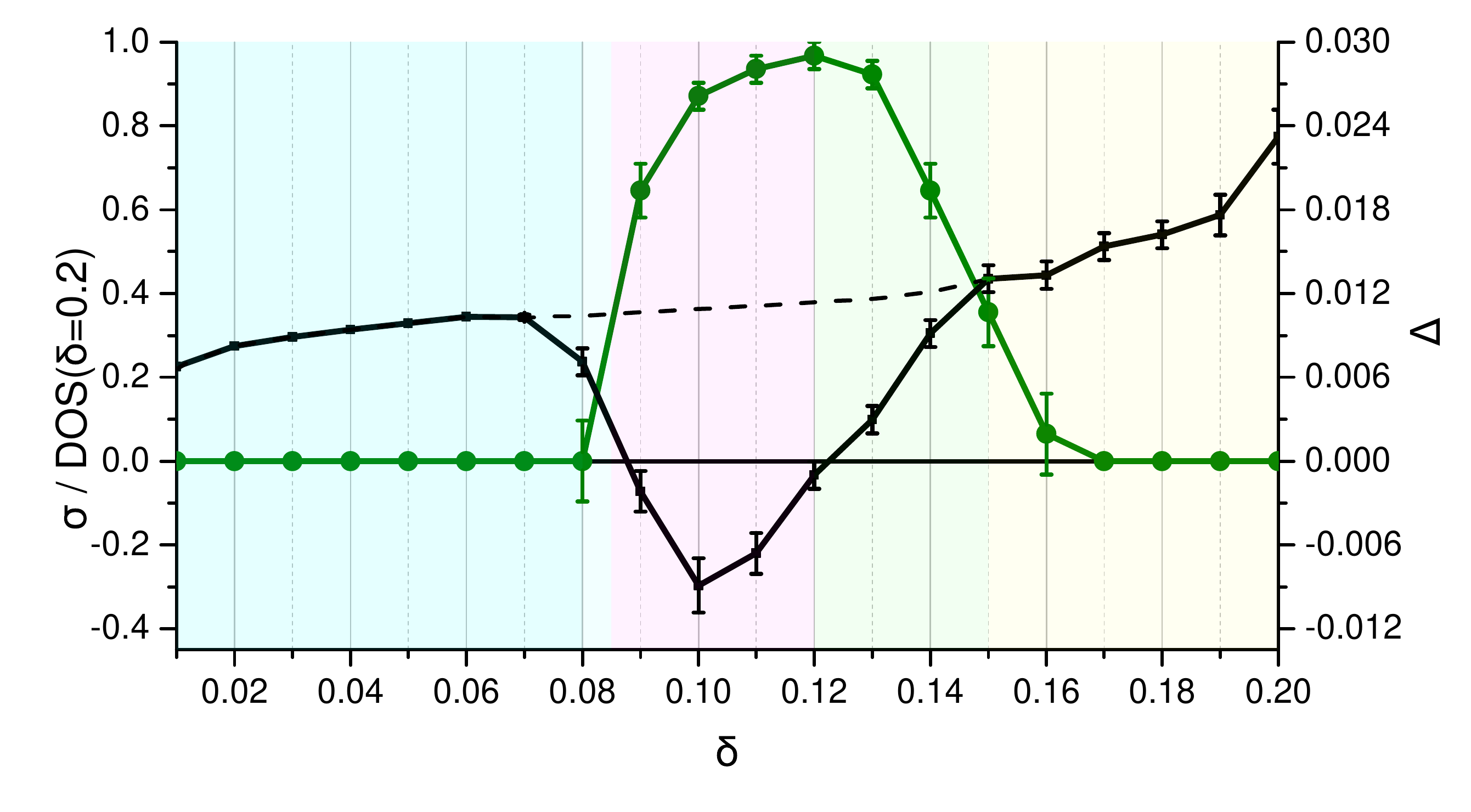}}
\end{minipage}

\caption{{\bf (a)} Panel displays the doping dependence of the density of states (DOS) at the Fermi level. Red and blue solid lines correspond to the DOS associated with the electron/hole carriers, respectively.
{\bf (b)} The black line shows the doping dependence of the calculated $\sigma$ coefficient (left vertical axis).
The green line shows the doping dependence of $\Delta$ (right vertical axis).
The dashed black line in panels {\bf (a)} and {\bf (b)} correspond to the "pure" PG phase without FSR.}
\label{Fig4}
\end{figure}

The electron-electron interaction opens up the gap in the Bragg reflection points (Fig.\ref{Fig3}(b)) whose value varies linearly with $U$(Fig.\ref{Fig3}(f)). This is present in the doping range $0.09<\delta<0.16$ (Fig.\ref{Fig3}(e)) which implies a splitting of the single band into the conduction and valence bands.
A direct calculation of the sign of the inverse effective mass by computing the second derivative of the energy dispersion ($m^{-1}_{\mathrm{eff}}=\partial^2E(\mathbf{k})/\partial\mathbf{k}^2$) allows us to reproduce the key features of the model in the wide doping range.
The results presented in the Fig.\ref{Fig4}(a) agree with the observation that at $\delta<0.09$ the conductivity is fully hole-like.
Qualitatively the same behavior is obtained for dopings $\delta>0.15$ where the carriers are also predominantly hole-like.
This characterized the "pure" PG phase.
However, the FSR which starts at $\delta\approx0.09$ and ends at $\delta\approx0.15$ leads to a drastic drop of the hole concentration at the Fermi level and to the emergence of the electron-like carriers.
Exactly in this region the FS is characterized by the presence of the Bragg gap and the emergence of an "average sign" of carriers ($\sigma=\int_{BZ}A(\mathbf{k},\varepsilon_{f})\mathrm{sign}(m_{\mathrm{eff}})d\mathbf{k}$). This is sharp contrast with the features expected in the "pure" PG phase (Fig.\ref{Fig4}(b)).
Moreover the total spectral weight at the Fermi level undergoes a rapid drop (Fig.\ref{Fig4}(a)) in agreement with heat capacity measurements\cite{Michon2018,Proust2018}.

Our approach successfully reveals some subtle details of the FSR resulting from a specific structure of the FS in the CDW phase.
While there are arguments for both the presence, in the CDW phase, of the electron pockets either alone or in the company with the two hole-like pockets, we show that those two competing situations become indeed manifest in different doping regimes.
More specifically, the obtained FSR reveals that, with an increasing $\delta$, the electron-like pocket in the nodal region emerges first at $\delta \approx 0.09$ and only after that at a slightly larger $\delta\approx 0.12$, two small hole-like pockets make their appearance in the FS in addition to the already existent electron-like pockets.
This explains why the Seebeck coefficient measurements \cite{DL2015} indicate that the {\it fully electron-like carriers} exist only within a thin region around $\delta\approx 0.10$.
Such a behavior implies that the electron-like pocket may exist either alone or in the company of the additional hole-like pockets which also contribute significantly to the magnetoresistance\cite{Vignolle2012}.

\paragraph{Conclusion.}

A new phenomenological model proposed in this Letter provides a fresh look at the unusual behavior exhibited by the FS in underdoped cuprates.
Although the translation symmetry breaking is introduced phenomenologically, electron-electron interaction is considered within the $t-J$ model of strongly correlated electrons thereby providing a full microscopic treatment of the short-range electron correlations.
In the CDW phase the closed electron-like pockets are indeed produced by the translational symmetry breaking, but the distribution of the spectral weight corresponds to the arc-like FS.
Due to the Bragg reflections on the boundaries of the BZ the effectively closed orbits of the quasiparticles emerge 
giving rise to quantum oscillations in a magnetic field.
Away from the CDW phase, the Bragg reflection is no longer possible.
As a result the "pure" PG phase that displays the disconnected FS arcs immediately sets in.
The resulting FS features and the ensuing charge carrier properties are uncovered in detail across the whole PG doping range in a very good agreement with experiment.

\end{document}